\documentclass[11pt,a4paper]{article}
\pdfoutput=1
\usepackage[utf8]{inputenc}
\usepackage[T1]{fontenc}
\usepackage{lmodern}
\pdfmapfile{+lm.map}
\usepackage{amsmath,amssymb,mathtools}
\usepackage[margin=1in]{geometry}
\usepackage{microtype}
\usepackage{booktabs,longtable,array,calc}
\usepackage{caption}
\usepackage{float}
\usepackage{graphicx}
\usepackage{xcolor}
\usepackage{listings}
\usepackage{newunicodechar}
\usepackage{hyperref}
\usepackage{bookmark}
\usepackage{xurl}
\hypersetup{colorlinks=true,linkcolor=blue!55!black,urlcolor=blue!55!black,pdftitle={Fixed-Point Scaffolding in the Clef Programming Language: Our Theoretical Grounding for Type-Preserving Compilation and Proof Inference},pdfauthor={Houston Haynes}}
\newcommand{\pandocbounded}[1]{\begingroup\setbox0\hbox{#1}\ifdim\wd0>\linewidth\resizebox{\linewidth}{!}{#1}\else#1\fi\endgroup}
\newsavebox{\paperdisplaybox}
\let\paperdisplayopen\[
\let\paperdisplayclose\]
\long\def\[#1\]{\begingroup
  \sbox{\paperdisplaybox}{$\displaystyle #1$}%
  \ifdim\wd\paperdisplaybox>1.25\linewidth
    \PackageError{paper}{Display requires source line breaks}{Split the display in Markdown.}%
  \fi
  \paperdisplayopen
  \ifdim\wd\paperdisplaybox>\linewidth
    \typeout{Paper display scaled to fit width: \the\wd\paperdisplaybox}%
    \resizebox{\linewidth}{!}{\usebox{\paperdisplaybox}}%
  \else\usebox{\paperdisplaybox}\fi
  \paperdisplayclose\endgroup}
\newunicodechar{α}{\ensuremath{\alpha}}
\newunicodechar{β}{\ensuremath{\beta}}
\newunicodechar{ε}{\ensuremath{\varepsilon}}
\newunicodechar{θ}{\ensuremath{\theta}}
\newunicodechar{κ}{\ensuremath{\kappa}}
\newunicodechar{Σ}{\ensuremath{\Sigma}}
\newunicodechar{⊗}{\ensuremath{\otimes}}
\newunicodechar{⊕}{\ensuremath{\oplus}}
\newunicodechar{↔}{\ensuremath{\leftrightarrow}}
\newunicodechar{¬}{\ensuremath{\neg}}
\newunicodechar{∧}{\ensuremath{\land}}
\newunicodechar{≤}{\ensuremath{\leq}}
\newunicodechar{×}{\ensuremath{\times}}
\newunicodechar{·}{\ensuremath{\cdot}}
\newunicodechar{†}{\ensuremath{\dagger}}
\newunicodechar{⟩}{\ensuremath{\rangle}}
\newunicodechar{²}{\ensuremath{{}^2}}
\newunicodechar{¹}{\ensuremath{{}^1}}
\newunicodechar{⁻}{\ensuremath{{}^-}}
\newunicodechar{ᵢ}{\ensuremath{{}_i}}
\newunicodechar{₀}{\ensuremath{{}_0}}
\newunicodechar{₁}{\ensuremath{{}_1}}
\newunicodechar{₂}{\ensuremath{{}_2}}
\newunicodechar{₊}{\ensuremath{{}_+}}
\begin{document}
\begin{center}

\textbf{Fixed-Point Scaffolding in the Clef Programming Language}

\emph{Our Theoretical Grounding for Type-Preserving Compilation and Proof Inference}

\textbf{Houston Haynes}\\
Braidpoint, Asheville, NC\\
\href{mailto:research@braidpoint.tech}{\nolinkurl{research@braidpoint.tech}}\\
June 2026

\end{center}

\begin{center}\rule{0.5\linewidth}{0.5pt}\end{center}

\section*{Abstract}\label{abstract}
\addcontentsline{toc}{section}{Abstract}

Native compilation changes how a program represents values and performs operations from processor to processor. In our view, proofs that justify source-level decisions must remain connected to those changes. We develop fixed-point scaffolding for this continuity in Clef and our Composer compiler. Our Program Semantic Graph retains types, joint constraints and proof dependencies. A thin MLIR middle end witnesses their settled consequences for native CPU, GPU, NPU and FPGA realization.

The account connects nanopass traversal with preservation contracts, structural theorems and adjoint interfaces between reasoning modes. Four verification tiers organize automatic inference and reusable domain and system proofs, with probabilistic relations admitted at Tier 4. Established lower-tier facts can directly support richer derivations. We state conditions for composing lowering contracts and assembling compatible program segments, then show how their dependencies govern differential compilation after an edit.

A worked foreign-call boundary follows source ranges through argument marshaling, the actual call and return conversion. Current FFI mechanisms bind checking evidence to those witnessed operations. A recent paper from Urschel provides elimination-growth bounds that supply a quantitative example. Reusable premises inform representation and error contracts that retain their sampling, arithmetic and exceptional-path assumptions. A further example admits value-indexed pairing through ordinary MLIR operations, connecting this scaffold to negative and fractional types.

Together these constructions explain how a thin middle end can carry the consequences of rich source proofs into unboxed computation graphs. Integrity and efficiency share the same retained evidence. It justifies native realizations and identifies which analyses, proofs and lowered fragments remain reusable during development.

\section*{1. Preserving the Reason for a Decision}\label{preserving-the-reason-for-a-decision}
\addcontentsline{toc}{section}{1. Preserving the Reason for a Decision}

A compiler makes decisions that a developer should be able to rely on: whether a value fits its representation, whether a buffer remains live through a call, and whether independently compiled pieces agree at their boundary. Those decisions become useful guarantees only when the emitted operations satisfy the conditions that justified them. An integer range established before lowering, for example, must constrain the argument actually passed to a foreign function and the conversion actually applied to its result.

Our Fidelity Framework makes that continuity part of compilation. Clef expresses the program. Our Program Semantic Graph (PSG) retains its identities, joint constraints and proof dependencies. Within Clef Compiler Services, Baker establishes and transforms the semantic structure. Alex witnesses its settled operations in portable MLIR, and Composer\textquotesingle s backends realize them for the selected target. A lowering boundary therefore carries two related products: executable structure and evidence connecting that structure to its source contract.

Fixed-point scaffolding names our organization of this work. A nanopass dispatcher handles nested scopes, local transformation rules establish preservation, and supported analyses settle the facts those rules consume. Each construction has its own justification. Their connection lets a compiler reuse an established result while making its application to a particular operation inspectable. Application developers work with ordinary Clef programs. The compiler supplies the covered proof applications and reports the premise that prevents a decision when one remains unresolved.

Three foundations give this architecture its formal direction. Ohori treats typed compilation as transformation between proof systems \hyperref[ref-ohori]{{[}1{]}}. Parametricity supplies theorems shared by every term in a specified type interpretation \hyperref[ref-reynolds]{{[}2{]}} \hyperref[ref-wadler]{{[}3{]}}, with Kennedy\textquotesingle s dimensional work giving a particularly useful instance \hyperref[ref-kennedy1996]{{[}4{]}} \hyperref[ref-kennedy2009]{{[}5{]}}. Adjoint logic supplies interfaces between reasoning disciplines, including the permissions to reuse or consume their assumptions \hyperref[ref-pruiksma]{{[}6{]}} \hyperref[ref-havarneanu]{{[}7{]}}. We connect these foundations through preservation contracts bound to the PSG\textquotesingle s actual operations, participants and target declarations.

The contribution is both formal and engineering-oriented. We state how local preservation contracts compose, distinguish stage compatibility from the assembly of separately analyzed segments, and specify how evidence crosses reasoning modes and representation boundaries. Our foreign-call example follows a value from its source range through C-compatible marshaling and back into a Clef result. Urschel\textquotesingle s recent elimination-growth bounds provide a complementary example of mathematical evidence entering this pipeline \hyperref[ref-urschel]{{[}8{]}}. The negative and fractional type case then tests the same scaffold against value pairing, resource use and inverse-operation contracts. Our thin MLIR middle end carries their established consequences into unboxed target operations, even where the target vocabulary has no corresponding source type.

Here, inference means the derivation of types, facts and proof applications during compilation. Machine-learning inference is one possible application of the resulting program.

Our companion papers establish the material this account must preserve. DTS/DMM develops dimensional identity and memory-lifetime reasoning, so lowering can rely on more than an erased machine type \hyperref[ref-dts]{{[}9{]}}. PHG retains joint relationships and their complete participant sets, which determine which operations and proofs must change together \hyperref[ref-phg]{{[}10{]}}. DBC develops the four verification tiers and their application to concurrent and distributed computation \hyperref[ref-dbc]{{[}11{]}}. FPS addresses the next question: how those established facts continue to justify a program as it is transformed, specialized and revised.

These relationships also govern efficiency. A proved transformation can reuse its general preservation theorem. An application checks the particular premises. An edit can invalidate the dependent region while leaving supported evidence elsewhere available. The same retained information that permits a justified native representation thus supplies the basis for differential compilation throughout the development loop.

\section*{2. Compatibility Across Stages and Segments}\label{compatibility-across-stages-and-segments}
\addcontentsline{toc}{section}{2. Compatibility Across Stages and Segments}

Consider a source value whose admitted range fits a particular signed representation. One stage establishes the range, another chooses the representation, and a later stage emits the arithmetic. The connection to preserve is the value\textquotesingle s interpretation at each stage. Copying a range label is useful only when it remains associated with the operation and operands that satisfy it.

\subsection*{2.1 Coherent translations}\label{coherent-translations}
\addcontentsline{toc}{subsection}{2.1 Coherent translations}

Let \(F(s)\) be the space of relevant semantic facts at compilation stage \(s\), and let \(D_{s,t}\) translate those facts along an admitted lowering from \(s\) to \(t\). For a stage order, the coherence requirements are

\[D_{s_1,s_2}\circ D_{s_0,s_1}=D_{s_0,s_2},
\qquad D_{s,s}=\mathrm{id}.\]

These are the identity and composition laws of a functor: a structure-preserving assignment of fact spaces and their translations to the stages and paths of compilation. The equality uses the selected semantic interpretation or proof equivalence. Different native instruction sequences can realize that same interpretation. When multiple lowering routes connect two stages, their coherence must be established before a poset, with one arrow per comparable pair, can represent those routes.

A compatible family chooses \(x_s\in F(s)\) at each stage such that

\[D_{s,t}(x_s)=x_t\quad\text{whenever }s\le t.\]

This is a global section of the diagram. On a finite poset, checking a supplied family on the cover relations---the immediate edges in its Hasse diagram---suffices, once the translations satisfy the composition laws. Every comparable pair factors through a chain of these edges. Substituting the checked equalities along the chain proves compatibility for that pair. This reduces the inventory of checks to the immediate boundaries, with the cost of each check determined by its fact space and comparison procedure.

Our compilation-sheaf terminology refers to this organization of compatible local information. It separates checking an existing family from finding one that extends independently obtained local solutions. The latter is an assembly problem, developed below through an explicit relational interface. A located contradiction or incompatible boundary is already a useful diagnostic. It identifies which program facts and translations need attention.

\subsection*{2.2 Joint facts and representation changes}\label{joint-facts-and-representation-changes}
\addcontentsline{toc}{subsection}{2.2 Joint facts and representation changes}

The facts being connected have different carriers. Physical dimensions over fixed base units use \(\mathbb Z^n\). Escape classifications and possible blade supports use ordered domains. Value ranges use intervals or richer predicates. Probability distributions carry weights as well as support. Our PSG connects these disciplines through shared program identities and declared transfer rules, while retaining the algebra appropriate to each.

Tarau\textquotesingle s work on bijective encodings provides a useful discipline for a change of representation: specify its carrier and check its round-trip law \hyperref[ref-tarau]{{[}12{]}}. For a codec into an admitted representation \(R_A\),

\[\mathrm{dec}_A\circ\mathrm{enc}_A=\mathrm{id}_A.\]

The decoder operates on valid encodings, and a further law states how each relevant operation is preserved. For example, encoding a value exactly and adding its encoded bits are separate obligations. A rounded conversion instead needs a stated relation or error bound. Our FFI example uses both exact conversion and representation coverage to connect this distinction to an actual call.

Joint constraints retain everyone a decision depends on. A shared-capacity obligation includes all buffers competing for the capacity. A foreign call includes its ordered argument occurrences, returned value and resource relationships. Auxiliary nodes in an ordinary graph can encode such relations. Our hypergraph expresses them directly and retains their participant identities, order and multiplicity where those affect the computation. That inventory supplies the premises against which a boundary is checked and the dependency set used when a program changes.

\subsection*{2.3 Assembling compatible program segments}\label{assembling-compatible-program-segments}
\addcontentsline{toc}{subsection}{2.3 Assembling compatible program segments}

Parallel compilation raises an additional question: can results established separately for several segments be assembled into a compatible result? Each segment may have several admissible interface assignments. Agreement about which individual values are possible does not always mean that the segments can choose those values together.

We make one sufficient condition precise using the relational setting of Beeri, Fagin, Maier and Yannakakis \hyperref[ref-beeri]{{[}13{]}}. Take a finite nonempty family of segments. For segment \(i\), let \(I_i\) be its finite set of interface identities and let \(R_i\) be a nonempty relation of admissible assignments to \(I_i\). Each identity has the same value domain wherever it occurs. A \emph{join tree} arranges the segments in a tree so that all segments containing any given identity form a connected subtree. This running-intersection condition gives each shared identity one connected route through the interfaces.

\textbf{Proposition 1 (relational segment assembly).} Suppose the interfaces have a join tree and adjacent relations agree on their shared projection:

\[\pi_{I_i\cap I_j}(R_i)=\pi_{I_i\cap I_j}(R_j).\]

Then the natural join \(R=\Join_i R_i\), the assignments satisfying every segment relation, is nonempty and satisfies \(\pi_{I_i}(R)=R_i\) for every \(i\).

\emph{Proof.} Root the tree at any segment and choose any tuple in its relation. Equal separator projections allow a matching tuple to be chosen in each child. Continue outward. Running intersection ensures that a child\textquotesingle s identities shared with an already visited segment also occur in its parent, so each extension agrees with all earlier choices. The assembled tuple belongs to every relation. Rooting at any \(i\) and starting from any tuple in \(R_i\) proves \(R_i\subseteq\pi_{I_i}(R)\); the reverse inclusion follows from the definition of the join. \(\square\)

For contrast, take Boolean interfaces \(\{x,y\}\), \(\{y,z\}\) and \(\{z,x\}\) with relations \(x=y\), \(y=z\) and \(z\ne x\). Each shared one-variable projection is \(\{0,1\}\), yet no assignment satisfies all three relations. Their cyclic interface pattern supplies no join tree. This small example shows why the compiler needs a structural certificate or a joint proof of compatibility before publishing independently settled results.

Our compiler use of Proposition 1 requires the relations to cover the actual shared values, effects and resources, and the separator projections to describe the same contracts. A containment tree or a scheduling certificate alone supplies neither correspondence. The connected-subtree certificates developed in PHG offer a source of structural evidence; the segment interface relation supplies the semantic evidence. With both established, fold-in can assemble independently checked segments. A remaining coupled component is settled through its joint obligations before its boundary is published. The theorem organizes that assembly. Constructing and comparing its relations remain part of the compiler\textquotesingle s cost.

\section*{3. Proof Inference Through Four Tiers}\label{proof-inference-through-four-tiers}
\addcontentsline{toc}{section}{3. Proof Inference Through Four Tiers}

For an application developer, the useful unit of verification is an ordinary program operation whose requirements the compiler can establish. Some requirements follow from types, some from arithmetic, and others from a reusable theorem about several participants. Our tiers organize these sources of evidence and the procedures that establish their premises \hyperref[ref-dbc]{{[}11{]}}.

{\def\LTcaptype{none} 
\begin{longtable}[]{@{}
  >{\raggedright\arraybackslash}p{(\linewidth - 6\tabcolsep) * \real{0.2500}}
  >{\raggedright\arraybackslash}p{(\linewidth - 6\tabcolsep) * \real{0.2500}}
  >{\raggedright\arraybackslash}p{(\linewidth - 6\tabcolsep) * \real{0.2500}}
  >{\raggedright\arraybackslash}p{(\linewidth - 6\tabcolsep) * \real{0.2500}}@{}}
\toprule\noalign{}
\begin{minipage}[b]{\linewidth}\raggedright
Tier
\end{minipage} & \begin{minipage}[b]{\linewidth}\raggedright
Representative content
\end{minipage} & \begin{minipage}[b]{\linewidth}\raggedright
Method of establishment
\end{minipage} & \begin{minipage}[b]{\linewidth}\raggedright
Evidence carried forward
\end{minipage} \\
\midrule\noalign{}
\endhead
\bottomrule\noalign{}
\endlastfoot
\mbox{1} & Dimensions and declared structural relationships & Kinded inference, normalization and structural rules & Substitutions or derivations in the declared algebra \\
\mbox{2} & Local range, layout, arithmetic and access conditions & Specialized analysis or a supported solver fragment & The obligation, actual operands, premises and checking evidence \\
\mbox{3} & Domain, resource, protocol and distributed-system properties & Reusable theorems instantiated from program structure & The law, participant mapping and discharged use-site premises \\
\mbox{4} & Computational and probabilistic relations between executions & Derivations in an admitted relational logic & The relation, checked derivation and evidence for its leaves \\
\end{longtable}
}

Tiers 1--3 generate and discharge covered obligations without application-developer proof annotations. Library authors establish the general laws and their semantic bindings. The compiler supplies the program-specific instances. Tier 4 admits project annotations identifying the relation to establish, with the same library direction extending toward recognition of relational constructions from hypergraph structure. The tier labels describe how work is organized. Reasoning modes, introduced in §5, govern the logical interfaces through which its results compose.

The lower tiers deliberately favor tractable structural analyses and admitted arithmetic fragments. Kennedy\textquotesingle s dimensional inference supplies a principal-inference foundation for its specified calculus \hyperref[ref-kennedy1996]{{[}4{]}}. Integer exponents require integer-preserving solving and divisibility checks. Grade, support and lifetime facts use their own transfer rules. General quantifier-free integer and bitvector problems include NP-complete cases, so solver dispatch retains the identity and limits of the selected procedure. Cheap local evidence can become a premise of a more expensive proof without claiming that all higher-tier reasoning has the same complexity.

\subsection*{3.1 From a source property to its point of use}\label{from-a-source-property-to-its-point-of-use}
\addcontentsline{toc}{subsection}{3.1 From a source property to its point of use}

Our PSG retains type identity, value ranges, target declarations, layout facts and outstanding obligations with the operations they describe. Typing rules and abstract interpretation establish some facts directly. Hoare rules or weakest-precondition reasoning generate others. A sound sufficient precondition gives the compiler an actionable obligation even when an exact weakest precondition is unnecessary or costly.

BAREWire\textquotesingle s layout and boundary contracts enter this same mechanism before witnessing. The realized offsets, extents, alignment and accesses must correspond to the established contract. Local memory, IPC and network communication use this discipline, with explicit encoding or conversion relations where local and boundary representations differ. The proof remains attached to the argument or value occurrence that crosses the boundary. Flat closures and code/environment pairs make that association especially important. The call must use the environment belonging to the selected function.

A lowering rule then relates source and target states under those premises. An unbounded integer addition and an overflowing machine addition illustrate why that relation matters. The source range must cover the actual target intermediate, including its signedness. Each affected edge consumes a preservation theorem covering the transformation or validation evidence about the particular emitted result. External evidence records report these graph-carried decisions and their checks. Our PSG remains the semantic owner.

\subsection*{3.2 Domain and system laws as reusable proof material}\label{domain-and-system-laws-as-reusable-proof-material}
\addcontentsline{toc}{subsection}{3.2 Domain and system laws as reusable proof material}

A reusable law packages a result with the conditions under which it can be applied. For a bounded accumulation, those conditions include the admitted inputs, accumulator capacity and conversion rule. For a communication protocol, they include the participants, resources transferred and allowed message histories. The compiler recognizes the supported construction, instantiates the law and checks those conditions against the actual graph.

Iris supplies a mechanized foundation for ownership and invariant reasoning \hyperref[ref-iris]{{[}14{]}}. Actris develops dependent separation protocols for reliable FIFO communication \hyperref[ref-actris]{{[}15{]}}. Aneris addresses distributed execution \hyperref[ref-aneris]{{[}16{]}}, and RCLib uses that foundation for reliable communication, RPC and replicated storage over a network that can lose, duplicate and reorder messages \hyperref[ref-rclib]{{[}17{]}}. Our bindings connect their applicable judgments to Clef operations and BAREWire boundaries. These laws can serve ordinary communicating components as well as explicit actors. Their use follows from the program\textquotesingle s relationships.

The preservation duty is concrete. If a law transfers exclusive access to a buffer, lowering must retain the buffer identity and the events that transfer and end that permission. If a distributed law accepts a contribution once, the realized receive and state-update operations must implement that acceptance rule. Arithmetic, ownership and protocol evidence therefore meet at the same operation, where the compiler checks that their premises are compatible.

Relational proofs connect this account to source-to-target behavior and probability. Probabilistic relational Hoare logic relates distributions of executions, often by a coupling \hyperref[ref-barthe]{{[}18{]}}. A compiler-relational judgment relates observations under source and target semantics. Rocq provides a setting in which the applicable rules, their soundness and reusable derivations can be checked. Arranz-Olmos and colleagues give a concrete precedent in Jasmin. Their work connects a relational compiler argument to a probabilistic security interpretation, with the revised result addressing ML-KEM under its stated adversary assumptions \hyperref[ref-jasmin]{{[}19{]}}.

Our typed-quotation design gives library authors a Clef-facing form for these declarations. A quotation carries the proposition, parameters, premises and accepted justification. Elaboration binds the instance to graph identities and dispatches its leaves. Covered library applications become part of ordinary compilation. The editor can expose the resulting proof and dependencies for inspection without making acknowledgment a condition for using a Tier 3 law.

\subsection*{3.3 Evidence, deferral and trust}\label{evidence-deferral-and-trust}
\addcontentsline{toc}{subsection}{3.3 Evidence, deferral and trust}

A conservative range, a contradictory dimensional equation and an unfinished proof search require different responses. A stronger applicable law can refine the range. A contradiction identifies a premise or program decision to correct. An unresolved obligation retains its identity and assumptions during editing until the compilation boundary that requires a decision. Representation selection proceeds when its coverage obligations are established; an empty admissible set produces a located error.

Probabilistic evidence has similarly explicit meaning. Logical constraints determine admissible values. A probability distribution assigns weights to those values under specified assumptions. A theorem about that distribution can yield a bound and a failure budget, which subsequent rules must retain. This distinction is central both to ADM\textquotesingle s adaptive computations \hyperref[ref-adm]{{[}20{]}} and to the elimination example in §7.

The checking basis travels with the claim. A solver verdict relies on the solver and its encoding. Independently replayed proof output transfers the relevant reliance to the checker and its accepted rules, while source interpretation and boundary mappings remain explicit dependencies. An unsatisfiable core locates a contradictory subset; a replayable proof establishes its inconsistency \hyperref[ref-cvc5]{{[}21{]}}. Our terminal evidence identifies the property, artifact, target, participants, assumptions, rule versions and checking basis so that a consumer can determine exactly which result is justified.

\section*{4. Nanopass Traversal and Compositional Preservation}\label{nanopass-traversal-and-compositional-preservation}
\addcontentsline{toc}{section}{4. Nanopass Traversal and Compositional Preservation}

The scaffold has an implementation shape that is familiar from ordinary compiler construction. A witness handles one operation, and a scope-bearing operation invokes the same dispatcher on its nested structure. The preservation argument follows that shape: local rules consume the evidence for their inputs, scope-bearing cases use the established contracts of their children, and the result retains the interpretation required by its consumer.

Several different constructions participate in this process. Keeping them distinct makes their proof duties visible.

{\def\LTcaptype{none} 
\begin{longtable}[]{@{}
  >{\raggedright\arraybackslash}p{(\linewidth - 4\tabcolsep) * \real{0.3333}}
  >{\raggedright\arraybackslash}p{(\linewidth - 4\tabcolsep) * \real{0.3333}}
  >{\raggedright\arraybackslash}p{(\linewidth - 4\tabcolsep) * \real{0.3333}}@{}}
\toprule\noalign{}
\begin{minipage}[b]{\linewidth}\raggedright
Construction
\end{minipage} & \begin{minipage}[b]{\linewidth}\raggedright
What it organizes
\end{minipage} & \begin{minipage}[b]{\linewidth}\raggedright
Required justification
\end{minipage} \\
\midrule\noalign{}
\endhead
\bottomrule\noalign{}
\endlastfoot
Nanopass dispatcher & Visits nested program scopes & Well-founded descent over the admitted structure \\
Pass composition & Connects successive transformations & Matching preservation contracts \\
Analysis fixed point & Settles mutually dependent facts & Sound transfer and justified convergence \\
Re-settlement after an edit & Retracts and rebuilds affected facts & Current support and dependency-aware invalidation \\
Self-referential source definitions & Describes a possibly recurring computation & Its evaluation semantics and the property claimed of observations \\
\end{longtable}
}

Alex\textquotesingle s current \lstinline!WitnessRegistry.fs! makes the dispatch construction explicit. Mutually dependent lazy bindings connect the registered witnesses to their shared dispatcher. A transfer owns the immutable registry and callback, with witness selection determined by the admitted target context. The scope witnesses receive it through calls of this form:

\begin{lstlisting}
LambdaWitness.createNanopass (fun () -> lazyCombinator.Value)
ControlFlowWitness.createNanopass (fun () -> lazyCombinator.Value)
MatchWitness.createNanopass (fun () -> lazyCombinator.Value)
\end{lstlisting}

The dispatcher selects a witness that handles the node and reports an unhandled node at that boundary. Lazy initialization ties the dispatch definition; descent processes the nested scopes. This is a fixed point of the traversal functional. The local cases and hypotheses for nested scopes then provide the places at which preservation evidence composes.

Our current ownership boundary keeps that witnessing layer thin. CCS/Baker settles continuation segments, captures, frame layout, argument adaptations and their semantic obligations in the PSG. Alex composes their portable MLIR operations. Composer\textquotesingle s selected backend supplies the target-specific realization and its required checks. A new target therefore adds realizations of established contracts, while the source analyses and shared proof material retain their owners.

\subsection*{4.1 Focus, Context and Segmented Witnessing}\label{focus-context-and-segmented-witnessing}
\addcontentsline{toc}{subsection}{4.1 Focus, Context and Segmented Witnessing}

A witness must reach the correct occurrence of an operation with its surrounding context intact. Huet\textquotesingle s zipper supplies the underlying navigation idea: a focused subtree is paired with the enclosing path needed to recover its place in the larger structure \hyperref[ref-huet]{{[}22{]}}. Our PSG traversal retains the focus, path and graph so that a witness consumes settled facts at that occurrence. Shared code or an equal node shape can occur in different environments. The context identifies which formation and receiving contract apply here.

This organization gives the fixed-point combinator a precise operational role. A local witnessing pattern consumes the settled operation and context, emits its physical consequences, and invokes the dispatcher for the admitted nested scopes. A flat closure, for example, composes patterns for its function value, environment and call while retaining their source-established relationship. The zipper supplies access to those facts; the pattern\textquotesingle s contract supplies the justification for its output.

Segmented compilation uses this mechanism within scopes authorized by the semantic analysis. Before witnessing begins, the compiler identifies the current occurrences, supporting premises, receiving contracts and replacement boundary for each segment. The zipper then navigates that admitted scope. Independent segments may be witnessed separately when their dependency and effect contracts permit it. Recombination checks their interfaces and current revision before accepting the combined artifact.

This separates three responsibilities that cooperate in differential compilation: dependency analysis determines the affected scope, zipper navigation visits its actual occurrences, and compositional preservation establishes the replacement\textquotesingle s meaning. A segment that crosses an unresolved shared obligation expands or waits for that obligation to settle. Once its boundary is established, the thin middle end can lower the segment using the same witnessing patterns as a larger compilation.

\subsection*{4.2 From local rules to a preservation contract}\label{from-local-rules-to-a-preservation-contract}
\addcontentsline{toc}{subsection}{4.2 From local rules to a preservation contract}

Ohori\textquotesingle s proof theory gives this arrangement a precise antecedent \hyperref[ref-ohori]{{[}1{]}}. In its terminating logical fragment, proofs describe typed code and cut reductions describe execution. A proof transformer is a partial proof with a hole on its major-premise path. Filling that hole with a proof of the matching sequent yields another proof:

\[\frac{\mathcal C[\,]:\Delta_1\Rightarrow\Delta_2\qquad
        \mathcal D:\Delta_2\vdash_c A}
       {\mathcal C[\mathcal D]:\Delta_1\vdash_c A}.\]

The common conclusion and intermediate context are the composition interface. Ohori\textquotesingle s extension to jumps and loops establishes type soundness for the extended transition system. This supplies the relevant form of preservation when execution may continue indefinitely. His account motivates compiler stages whose transformations carry explicit semantic contracts.

For Clef, let \(p\) denote a program artifact, \(a\) its retained facts, \(L_{s,t}\) a lowering and \(D_{s,t}\) the corresponding fact translation. We use the contract

\[\mathrm{Valid}_s(p,a)\land H_{s,t}(p,a)
\Longrightarrow
\mathrm{Valid}_t(L_{s,t}(p),D_{s,t}(a)).\]

Here \(\mathrm{Valid}\) states the property and interpretation of the artifact, and \(H\) states the transformation\textquotesingle s side conditions. Those conditions include the premises for dimension substitution, representation selection, resource transfer and the permitted observations of effects. For an arithmetic operation, validity includes the relation between its mathematical value and its machine realization.

\textbf{Proposition 2 (successive preservation).} Suppose \(s\to t\) and \(t\to u\) satisfy this contract. Write \(p_t=L_{s,t}(p)\) and \(a_t=D_{s,t}(a)\). Then

\[\begin{aligned}
&\mathrm{Valid}_s(p,a)\land H_{s,t}(p,a)\land H_{t,u}(p_t,a_t)\\
&\qquad\Longrightarrow
\mathrm{Valid}_u(L_{t,u}(p_t),D_{t,u}(a_t)).
\end{aligned}\]

\emph{Proof.} The first contract supplies \(\mathrm{Valid}_t(p_t,a_t)\). This and the second side condition supply the premises of the second contract. Its conclusion is the required validity at \(u\). \(\square\)

The force of this elementary composition is in its interfaces. A proof of the first step must supply the interpretation consumed by the second. A subsequent assumption remains an obligation until justified. With coherent transports, the result agrees with the composite translation of §2.1. The nanopass traversal uses the same rule at each local case, with induction over the finite visited structure and the induction hypotheses for its nested scopes. The runtime program can contain loops even though the compiler visits a finite representation of them.

A general pass theorem is thus reusable proof material. Its use site supplies actual premises and an association between theorem, input, output and artifact revision. Per-instance validation can establish a particular boundary relation when that is the admitted checking route. Together these routes turn local preservation arguments into an inspectable compilation chain.

The geometric optimizer proposed in ADM \hyperref[ref-adm]{{[}20{]}} gives this contract another application, using Karwowski, Nielsen, and Staton \hyperref[ref-karwowski-geometry-2026]{{[}23{]}} for its probabilistic interpretation. Here \(a\) must retain the declared observation, sampling law, parameter correspondence, output metric and analytic premises. A pass claiming exact geometric preservation must establish the corresponding metric relation as well as its behavioral contract. Under a smooth parameter correspondence \(r:X_s\to X_t\), the geometric equality is \(G_s=r^*G_t\). A bounded or contracting relation instead supports only consumers whose specification admits that relation. Metric equality by itself supplies neither equality of traces nor an inverse for the transformed computation. DBC \hyperref[ref-dbc]{{[}11{]}} develops the admission obligations; ADM gives the restricted natural-gradient construction.

This choice matters when operational details are removed. A pass may preserve the output law while changing the trace used by a geometric observer. If the selected optimizer consumes that geometry, the pass needs its geometric preservation evidence or a revised optimizer contract. Keeping the original geometry as static evidence is another possible design: it describes a retained source model and must be identified as such. It does not automatically become the geometry of the emitted program\textquotesingle s execution. A target evaluator for that source model still needs a justified relation to the retained metric computation. The source model, executable observer and evidence therefore remain associated through their explicit correspondence, including any numerical approximation.

The same preservation question reaches a foreign-function boundary. A callable ABI and a result-range contract do not establish that equal arguments determine the same result: a foreign operation may depend on mutable state, random choices or callbacks. Our proposed geometric interface therefore binds each foreign application to its admitted relation, effect context and validity premises. A sampler additionally needs its probability law and the observation contract consumed by the optimizer; derivatives concern the declared mathematical model, with a separate error relation for its machine realization. Restoring an actual sampled occurrence requires the recovery evidence described in our Negative and Fractional Type treatment. This extends the existing authority chain: the Farscape binding generator declares external contracts, our Baker component within Clef Compiler Services settles their use, the Program Semantic Graph retains application evidence, and our Bozzetto daemon coordinates the work and exposes its evidence. Native witnessing and consumption must preserve those distinctions before a geometric consumer can reuse them. The geometry adds a consumer of the boundary contract, as opposed to granting purity or reversibility to an imported symbol.

\subsection*{4.3 Static facts and unboxed realization}\label{static-facts-and-unboxed-realization}
\addcontentsline{toc}{subsection}{4.3 Static facts and unboxed realization}

An engineer needs the effects of a proof at runtime, not necessarily its data structures. Harper, Mitchell and Moggi\textquotesingle s phase distinction separates static and dynamic components of a module system \hyperref[ref-phase]{{[}24{]}}. In our compiler account, a proof can justify a representation or operation and remain available to compilation without becoming a field in every runtime value.

A measured scalar illustrates the distinction. Once its dimensional, range and boundary obligations are settled, its runtime representation can be an unboxed scalar. The compiler\textquotesingle s preservation relation still binds that scalar to the measured argument or result. A machine signature such as \lstinline!f64 -> f64! is the realized shape. The graph and evidence retain which source value and operation it realizes.

The same principle applies to a closure. Its finite capture list determines which fields require layout, but the lifetime proof also follows storage reachable through those fields. An immutable reference can point to mutable storage, and a deferred computation can memoize a shared result. Our placement and generalization rules preserve those identities while the backend chooses the established physical layout.

For arithmetic, erasure preserves the conditions of the selected regime. Exact quire accumulation requires coverage of all intermediate sums. A transformation that partitions a reduction carries those premises into the workers and the combination step. If a boundary rounds a partial result, the receiving proof uses that rounded-result contract. This lets the compiler select efficient native operations while keeping the particular equality or error bound that justified them.

CPU, GPU, NPU and FPGA routes can consequently use different layouts and operation sequences. Each consumes the same source contracts through its own admitted realization. Our thin middle end carries their settled consequences. The backend\textquotesingle s preservation duties connect those consequences to executable operations on the selected hardware.

\section*{5. Proof Transport Between Reasoning Modes}\label{proof-transport-between-reasoning-modes}
\addcontentsline{toc}{section}{5. Proof Transport Between Reasoning Modes}

A local range fact can help prove a property of an entire computation. For example, evidence that an index lies inside a buffer can become a premise of a relational proof comparing two implementations of a foreign-call wrapper. Its usefulness depends on retaining the same index, buffer and interpretation as it moves into the larger argument. Ownership of that buffer also has to follow the rules governing exclusive access.

Adjoint logic provides the formal discipline we draw on for such interfaces. In Aram Hăvărneanu\textquotesingle s \emph{Classical Adjoint Logic}, the mode order specifies permitted dependencies, while the structural signature determines whether assumptions may be discarded or reused \hyperref[ref-havarneanu]{{[}7{]}}. These are the logical rules behind distinctions such as a reusable bound on an immutable value and a permission to consume or transfer a resource. Earlier adjoint-logic and graded-modality work supplies related foundations \hyperref[ref-pruiksma]{{[}6{]}} \hyperref[ref-eades]{{[}25{]}}.

Hăvărneanu\textquotesingle s Theorem 3.18 establishes the shift adjunction in his oriented calculus; Theorem 4.20 gives the corresponding construction within the declared compositional interfaces of the ambient account. For an adjunction \(F\dashv G\), the unit and counit have types

\[\eta:\mathrm{Id}\Rightarrow GF,\qquad
  \varepsilon:FG\Rightarrow\mathrm{Id},\]

and satisfy the triangle identities

\[\varepsilon_F\circ F\eta=\mathrm{id}_F,\qquad
  G\varepsilon\circ\eta_G=\mathrm{id}_G.\]

These laws govern the specified proof transformations and their contexts. They permit coherent compositions through the interface. A shift may add structure or expose only a selected part of a judgment. Extracting the support of a probability distribution, for example, retains admissible values while the weights belong to the richer interpretation. An inverse round trip requires the additional conditions of an equivalence.

\subsection*{5.1 Reusing a premise directly}\label{reusing-a-premise-directly}
\addcontentsline{toc}{subsection}{5.1 Reusing a premise directly}

Our tiers organize proof work; the modes govern the use of assumptions. Their connection is established by assigning concrete judgments and resource rules to admitted interfaces. Consequently, a Tier 2 arithmetic result can supply a Tier 4 proof leaf directly. It need not pass through a Tier 3 derivation that restates the same result.

Suppose \(\pi\) establishes \(0\le i<n\) for the actual index and buffer length at a call. A relational rule comparing two safe loads can consume \(\pi\) together with its ownership and representation premises. The interface records the original arithmetic judgment, its interpretation in the receiving logic and the permissions governing each resource. Reusing \(\pi\) shares an established derivation while checking its applicability to this load. A changed buffer, index association or interpretation changes the dependencies of that application.

This is how the decidable lower tiers pay down the work of richer proofs. A library author establishes the translation and relational rule once. Applications instantiate their remaining conditions. Reconstruction, premise checking and revalidation remain costs of the application, but the higher-tier proof can begin with already justified facts about this program.

\subsection*{5.2 Coherence between proof transport and lowering}\label{coherence-between-proof-transport-and-lowering}
\addcontentsline{toc}{subsection}{5.2 Coherence between proof transport and lowering}

There are now two directions of transformation. Lowering changes the program representation. Mode transport changes the judgment in which a fact is used. At their intersection, our compiler must show that both routes concern the same realized operation and property.

Write \(C_m(p)\) for evidence about artifact \(p\) in reasoning mode \(m\). Let \(V_p:C_m(p)\to C_n(p)\) be an admitted mode transport, and let \(T^m_L:C_m(p)\to C_m(Lp)\) be the evidence transport for a lowering \(L\). On evidence and side conditions for which both routes are admitted, the required correspondence is

\[T^n_L\circ V_p\ \simeq\ V_{Lp}\circ T^m_L.\]

The equivalence \(\simeq\) is the chosen semantic or proof equivalence. Both routes justify the corresponding target property under matched assumptions. This is a coherence obligation of our compiler interpretation. Supplying it for the selected transports lets a fact proved before lowering serve the target judgment with its participants and resource conditions intact.

\includegraphics[width=1\linewidth,height=\textheight,keepaspectratio]{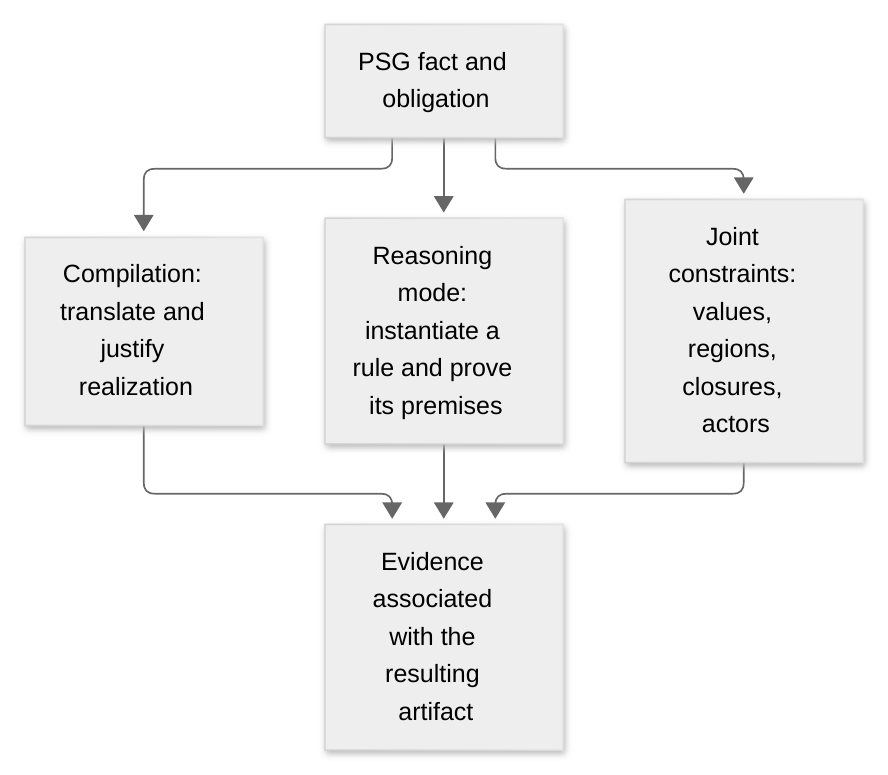}

The diagram gathers these duties at the evidence associated with an artifact. A mathematical law supplies its general result, the program supplies actual participants and premises, and lowering supplies their realized correspondence. The FFI example in §8 follows all three through one call.

\subsection*{5.3 Stateful protocols and logical fixed points}\label{stateful-protocols-and-logical-fixed-points}
\addcontentsline{toc}{subsection}{5.3 Stateful protocols and logical fixed points}

Repeated service of requests adds a further kind of structure. Qian, Kavvos and Birkedal\textquotesingle s client-server sessions describe a server threading state through client interactions, using least and greatest fixed points to define the protocol modalities \hyperref[ref-qkb]{{[}26{]}}. This gives an actor-oriented interpretation a precise source of rules for stateful service. Ordinary communicating components can also use admitted protocol laws without adopting actor syntax.

The logical fixed points in those modalities describe protocol behavior. Our nanopass dispatcher visits its finite representation, and adjoint shifts govern evidence interfaces. A combined calculus needs the cut and composition results appropriate to that combination. The existing library workflow can instead use a proved protocol rule with explicitly discharged application premises. Order-independent handling requires a law about the state transformations themselves, and completion requires the applicable scheduling or progress assumptions. Keeping these judgments explicit lets the compiler preserve exactly the protocol property the application relies on.

\section*{6. Structural Theorems and Evaluation}\label{structural-theorems-and-evaluation}
\addcontentsline{toc}{section}{6. Structural Theorems and Evaluation}

The compiler can obtain some preservation premises directly from a program\textquotesingle s type. This is useful because a theorem shared by a type interpretation can apply at every covered use, reducing the need to rediscover the same property for individual functions.

Reynolds\textquotesingle{} relational account establishes that a well-typed term respects the relations assigned to its type variables \hyperref[ref-reynolds]{{[}2{]}}. Wadler\textquotesingle s familiar list example makes the result tangible \hyperref[ref-wadler]{{[}3{]}}:

\[g:\forall\alpha.\,[\alpha]\to[\alpha]
\quad\Longrightarrow\quad
\mathrm{map}(f)\circ g_A=g_B\circ\mathrm{map}(f).\]

In the total pure interpretation, \(f:A\to B\) can be any total function. Applying a uniform list transformation and then mapping the elements agrees with mapping first and applying the transformation at the new element type. The theorem supplies a reusable equality for that calculus and its primitives. Effects, partiality and observations of evaluation require the corresponding logical relation and admissibility conditions.

\subsection*{6.1 Dimensional identity and representation}\label{dimensional-identity-and-representation}
\addcontentsline{toc}{subsection}{6.1 Dimensional identity and representation}

Kennedy\textquotesingle s dimensional invariance gives Clef a useful instance of such uniformity \hyperref[ref-kennedy1996]{{[}4{]}} \hyperref[ref-kennedy2009]{{[}5{]}}. In additive exponent notation, a dimension-polymorphic multiplication has type

\[\mathrm{mul}:\forall d_1,d_2.\;
\mathbb R^{\langle d_1\rangle}\to
\mathbb R^{\langle d_2\rangle}\to
\mathbb R^{\langle d_1+d_2\rangle}.\]

A compatible interpretation of the primitives supplies their scaling behavior, while dimensional inference checks the relationship at each use. For a compiler transformation \(L\), let \(\alpha_s\) and \(\alpha_t\) extract the relevant interpretation at its source and target. Preservation requires

\[\alpha_t\circ L=D\circ\alpha_s.\]

The equation says that interpreting the lowered operation agrees with translating the source interpretation. Its proof binds the measured value to the actual target operands and result. A rounded realization adds its arithmetic and error contract. The compiler can then erase a runtime measure tag while retaining the justification for the operation it emits.

Other structural facts contribute through their own algebras. Homogeneous parity has a \(\mathbb Z_2\) character under the grade involution; a general multivector can have both even and odd components. Homogeneous grade is bounded by the algebra\textquotesingle s dimension, while multiplication can produce several grades. Blade-support transfer accounts for the metric and possible cancellation \hyperref[ref-phg]{{[}10{]}}. Keeping these distinctions lets the compiler combine dimensional and geometric information without assigning them identical inference rules.

\subsection*{6.2 Demand is part of the contract}\label{demand-is-part-of-the-contract}
\addcontentsline{toc}{subsection}{6.2 Demand is part of the contract}

Clef\textquotesingle s call-by-need semantics defers ordinary arguments, constructor payloads and record fields, sharing each result when demanded. Its explicit \lstinline!eager! form demands an outer value at an activated boundary. A rewrite therefore preserves not only returned values but also the required observations of demand, effects and termination.

This matters even for a self-referential structure that is easy to form. A stream with a constant head and a deferred reference to its tail can expose each finite prefix under its productive demand rule. A deferred head that demands itself again can fail to produce the first element, despite the surrounding constructor being available. The useful proof concerns the demands required for an admitted observation. Our graph account retains those dependencies with the group of mutually dependent definitions so that a scheduling or evaluation change has a precise preservation obligation.

Johann and Voigtländer\textquotesingle s treatment of Haskell\textquotesingle s \lstinline!seq! demonstrates why evaluation-sensitive primitives require a revised parametricity account \hyperref[ref-johann]{{[}27{]}}. Their result motivates a Clef-specific logical relation covering \lstinline!eager!, rather than importing the total-pure list equation unchanged. A library rewrite supplies its evaluation and effect premises along with the value equality. The compiler can then use the rule automatically in the fragment whose behavior it preserves.

These examples locate the practical role of structural theorems. They give the compiler reusable facts about a defined interpretation. Values, error bounds and relationships between executions add further premises through the same scaffold, as the next section illustrates with a quantitative theorem.

\section*{7. Quantitative Premises for Native Realization}\label{quantitative-premises-for-native-realization}
\addcontentsline{toc}{section}{7. Quantitative Premises for Native Realization}

A native linear solve needs a representation that covers its working values and an error contract appropriate to the result. Urschel\textquotesingle s recent work supplies a useful class of mathematical premises for that decision \hyperref[ref-urschel]{{[}8{]}}. Our contribution here is their admission as reusable evidence and their preservation through the operations that consume them. For a no-pivot factorization \(B=LU\), with unit lower-triangular \(L\), he defines

\[\rho(B)=\max\left\{\|L\|_{\max},\frac{\|U\|_{\max}}{\|B\|_{\max}}\right\},\]

where \(\|\cdot\|_{\max}\) is the largest absolute entry. This definition controls the factors. A bound on every intermediate value requires the corresponding analysis of the chosen algorithm.

Theorem 1.2 in Urschel\textquotesingle s paper fixes any nonsingular \(A\in\mathbb R^{n\times n}\) and draws \(Q\in O(n)\) from Haar measure. For elimination without pivoting on \(QA\), it gives

\[\Pr\!\left[\rho(QA)>T(n,\delta)\right]<\delta,
\qquad T(n,\delta)=\frac{14n^{3/2}\sqrt{\log(en)}}{\delta},
\qquad 0<\delta<1.\]

The randomness belongs to the preconditioner, so the theorem can apply to a structured input matrix. In the chosen dimensionless coordinates, \((QA)x=Qb\) has the same exact solution as \(Ax=b\), and orthogonality preserves the spectral condition number: \(\kappa_2(QA)=\kappa_2(A)\). The contract records the scaling that establishes those coordinates and the bounds needed when forming both transformed inputs.

Suppose \(M\ge\|QA\|_{\max}\) is established. On the event \(\rho(QA)\le T(n,\delta)\),

\[\|L\|_{\max}\le T(n,\delta),\qquad
  \|U\|_{\max}\le T(n,\delta)M.\]

These are useful factor-range bounds. For example, the exact factors satisfy \((|L||U|)_{ij}\le nT(n,\delta)^2M\). A backward-error theorem involving the computed factors \(|\widehat L||\widehat U|\) needs an arithmetic bridge connecting them to the mathematical factors. The resulting error contract also covers triangular solves and the actual rounding model. Conditioning then connects backward error to solution accuracy. This gives the compiler a route from a stated failure budget to a numerical contract for the realized solve.

Urschel\textquotesingle s Theorem 1.4 concerns a different input law and algorithm. If \(G\) has independent standard Gaussian entries and \(\Pi\) is the specified partial-pivoting permutation, then for every \(\alpha>0\) there are constants \(C_\alpha,n_\alpha\) such that

\[\Pr\!\left[\rho(\Pi(G))>
\sqrt n\exp(C_\alpha\sqrt{\log n})\right]<n^{-\alpha},
\qquad n\ge n_\alpha.\]

Together with Corollary 1.5, this resolves the conjectured growth exponent \(1/2\). Its constants and dimension threshold remain part of an application. The Haar result supplies the explicit finite-dimensional alternative above. \hyperref[ref-urschel]{{[}8{]}}

A library instance records the matrix construction, sampler interpretation, factorization algorithm, scaling, arithmetic model and admitted error budget. A finite-machine sampler and transform require their own connection to the ideal theorem. Lowering preserves those dependencies and the behavior of any exceptional branch. Deterministic storage safety is established by capacity bounds or guards checked before the potentially overflowing operation, with an admitted fallback. The probabilistic budget describes the numerical event within that safe execution contract. Changes to the sampler, precision, matrix construction or pivoting rule renew the affected obligations while unchanged structural evidence remains reusable.

The compiler-facing theorem instance consequently has a concrete shape. Tier 1 establishes dimensions and the coordinates in which the matrix theorem applies. Tier 2 checks covered size, range and conversion premises. A Tier 3 library application supplies the growth result under its sampling and algorithm assumptions, and a Tier 4 argument can relate the implemented solve to the specified result distribution. A mode interface can reuse each established premise where it is needed. Lowering carries the resulting factor, error and exceptional-path contracts to the actual target operations.

This makes the efficiency consequence reviewable. A justified range can narrow storage requirements, and an accepted accuracy contract can guide precision selection. The compiler retains why that choice was admitted, including its failure budget and safe exceptional behavior. The gain is a principled selection process shared across applications. A particular speed or energy improvement depends on the chosen algorithm, target and resulting checks.

\section*{8. Foreign Calls as Preservation Boundaries}\label{foreign-calls-as-preservation-boundaries}
\addcontentsline{toc}{section}{8. Foreign Calls as Preservation Boundaries}

A foreign call brings the abstract preservation conditions into a familiar engineering task. A C or C++ library exposes an ABI, a calling convention and an execution contract. Our compiler must connect those declarations to Clef\textquotesingle s actual arguments, conversions, returned values and resource lifetimes. Clef\textquotesingle s native compilation path uses the foreign interface as an explicit marshaling boundary. Its internal representation is selected through our own source and target contracts.

Our Farscape binding-generator application draws directly on the ML tradition of metaprogramming for interoperation. Foreign interfaces become typed descriptions from which bindings are generated. Farscape derives Clef bindings from C headers while retaining the distinction between source types and foreign ABI representations. Those declarations enter the same constraint and proof pipeline as ordinary Clef code, so the boundary\textquotesingle s integrity begins with how the binding is expressed.

At each use of a generated binding, the Baker front-end component in Clef Compiler Services records ordered actual and formal participants, checks their coverage obligations and publishes the required adaptations. Alex witnesses those adaptations and the corresponding call. Composer binds the source checking result to the current graph, obligation inventory and witnessed operations before backend production. Our specification describes these responsibilities in its FFI boundary contract.

\subsection*{8.1 One call, two coverage obligations}\label{one-call-two-coverage-obligations}
\addcontentsline{toc}{subsection}{8.1 One call, two coverage obligations}

Consider a declared foreign signature, written here in C notation solely to expose its ABI:

\begin{lstlisting}
int16_t foreign_step(int32_t x);
\end{lstlisting}

Suppose the source argument has an established range \(A=[0,1023]\), its settled representation is signed eleven-bit, and the receiving Clef expression uses signed thirty-two-bit storage. Write

\[I_w=[-2^{w-1},\,2^{w-1}-1].\]

Argument coverage requires \(A\subseteq I_{32}\). Result coverage requires \(I_{16}\subseteq I_{32}\) when the full declared result range is admitted. The argument and result use different premises: the former follows the actual source value, while the latter follows the foreign result declaration and receiving representation.

For an admitted integer calling convention, the corresponding portable operation sequence would likely have a schematic MLIR form \hyperref[ref-mlir]{{[}28{]}} similar to this:

\begin{lstlisting}
func.func private @foreign_step(i32) -> i16
func.func @wrapped_step(%x: i11) -> i32 {
  %arg = arith.extsi %x : i11 to i32
  %raw = func.call @foreign_step(%arg) : (i32) -> i16
  %result = arith.extsi %raw : i16 to i32
  func.return %result : i32
}
\end{lstlisting}

The displayed widths instantiate the checked design. MLIR integer types themselves are signless here. The signed interpretation is supplied by the contract and the \lstinline!arith.extsi! operations. If \(\operatorname{val}_w\) reads a bit pattern as a signed \(w\)-bit integer, the two conversion laws are

\[\begin{aligned}
\operatorname{val}_{32}(\operatorname{extsi}_{11,32}(x))&=\operatorname{val}_{11}(x),\\
\operatorname{val}_{32}(\operatorname{extsi}_{16,32}(r))&=\operatorname{val}_{16}(r).
\end{aligned}\]

The first applies to the actual source argument; the second applies to the value returned by this call. Backend lowering then establishes the declared target ABI, including any register or calling-convention adaptation. Proposition 2 composes these conversion and call contracts under the foreign function\textquotesingle s declared execution behavior.

Changing the receiving representation to signed eight-bit makes the full-result inclusion false. The compiler then needs an established stronger result range, a specified checked conversion with its failure behavior, or a receiving representation covering the declared result. Likewise, replacing sign extension with zero extension changes the interpretation of negative results. These are located proof obligations concerning a particular boundary, so a passing range check on an unrelated value cannot justify the call.

\subsection*{8.2 What binds the proof to the operation}\label{what-binds-the-proof-to-the-operation}
\addcontentsline{toc}{subsection}{8.2 What binds the proof to the operation}

A type alone cannot distinguish two arguments with the same machine representation. Our boundary retains actual occurrence identities and their ordered association with formal parameters. Repeated arguments retain their multiplicity. The result adaptation refers to the SSA result emitted by that call. These details make a theorem instance about the argument or return value applicable to the operation that will execute.

Callable values add another relationship: code is paired with an environment and a receiving signature. A branch that selects a callable also selects the corresponding formation and lifetime support. Our current FFI contracts retain these dependencies, giving the compiler a basis for checking the chosen call and invalidating its evidence when that support changes. An equal signature supplies compatibility of shape; the retained formation supplies the particular callable being invoked.

Our closure representation makes that formation explicit. A callable consists of a portable function value and a separate flat environment \lstinline!memref!, represented as two SSA values at the witnessing boundary. The environment contains the captured fields. The code value remains separate. {\textquotedblleft{}Fat} {pointer\textquotedblright{}} is useful shorthand for this logical code/context pair, whose physical ABI follows the admitted target contract. Selecting a callable through a branch selects its code and environment together. For example, closures formed by \lstinline!makeScale 2! and \lstinline!makeScale 3! can share the multiplication function while carrying different captured factors. Choosing the first closure must preserve its factor of two. A compatible function signature alone cannot establish that association.

The capture rules determine what those fields mean. An immutable capture retains the established value or shared deferred identity. A reference preserves the referenced storage\textquotesingle s sharing. A mutable capture retains access to its live storage cell. Baker settles this formation and its lifetime, and Alex composes the witnessing patterns for the function, environment and call. The pair is consequently an unboxed realization of a particular source closure with the necessary layout and lifetime evidence.

A C callback needs a further declared adapter. An API may accept a function pointer and a separate user-data argument, but that convention is an external contract to which our pair must be mapped. The adapter establishes argument and result correspondence, context identity and the retention/release behavior required by the API. This is where the flat-closure account meets marshaling. A correct callable formation inside Clef is carried through an explicit foreign convention, with its resource conditions preserved.

Buffers and callbacks extend the same pattern to ownership. A borrowed buffer requires extent, alignment, lifetime and permitted-access premises through the call. A callback requires the lifetime of its environment and the allowed invocation behavior. Their proof terms can come from the resource and protocol libraries of §3.2. The external library\textquotesingle s admitted contract states the behavior assumed at that boundary. Marshaling establishes how our program enters and leaves it. This scopes the guarantee to properties whose premises the compiler and binding actually provide.

The implementation evidence therefore substantiates a central FPS claim. Preservation requires correspondence to actual operations, not just a collection of formulas that separately pass checking. The thin witnessing layer realizes settled semantic decisions, while the checking path retains their connection to arguments, resources and returned values. That same discipline supplies the route for admitting richer source constructions through ordinary MLIR operations.

\section*{9. Admitting Negative and Fractional Types}\label{admitting-negative-and-fractional-types}
\addcontentsline{toc}{section}{9. Admitting Negative and Fractional Types}

Negative and fractional types place a stronger demand on the same boundary. A library operation can carry a paired requirement for a particular value, a consumable resource or a checked reverse operation. These relationships must survive lowering even though MLIR has no built-in interpretation of the proposed source types. FPS supplies the representation and preservation interface. Our companion paper develops their operational core and applications \hyperref[ref-nft]{{[}29{]}}.

\subsection*{9.1 The information that crosses the boundary}\label{the-information-that-crosses-the-boundary}
\addcontentsline{toc}{subsection}{9.1 The information that crosses the boundary}

The companion paper draws on Chen and Sabry\textquotesingle s value-indexed calculus \hyperref[ref-chen]{{[}30{]}}. A fractional requirement identifies a distinguished value. Matching the ordinary type alone does not establish that the required value has been supplied. Its negative construction uses a directed evaluation interpretation. Those source meanings determine the success, failure, resource and evaluation behavior that a realization must preserve.

Compact duality expresses part of the pairing interface. With a consistent ordering convention, a dual \(A^*\) has maps

\[\eta_A:I\to A^*\otimes A,\qquad
  \varepsilon_A:A\otimes A^*\to I,\]

satisfying, with associators and unitors suppressed,

\[\begin{aligned}
(\varepsilon_A\otimes\mathrm{id}_A)\circ(\mathrm{id}_A\otimes\eta_A)&=\mathrm{id}_A,\\
(\mathrm{id}_{A^*}\otimes\varepsilon_A)\circ(\eta_A\otimes\mathrm{id}_{A^*})&=\mathrm{id}_{A^*}.
\end{aligned}\]

Here \(I\) is the monoidal unit. These snake identities express coherence of creating and consuming a matched pair. Compact duality retains the paired structure and does not identify \(A\otimes A^*\) with \(I\). The companion paper develops this construction and its illustrated identities. An inverse of a particular computation additionally requires that computation\textquotesingle s inverse equations, and a value-indexed match retains its specific equality judgment. Our compiler interface therefore records which of these properties a library operation actually promises.

The geometric application in §4.2 adds a distinct interface. A derivative transports tangent vectors; its dual transports covectors. For finite-dimensional tangent spaces with nondegenerate metrics, a metric adjoint can be defined by the corresponding pairing equation. Degenerate metrics require the kernel compatibility needed for that equation to have solutions, followed by an admitted quotient or a specified choice of representative. None of these constructions alone supplies the compact unit and counit above, a value-indexed match, or an executable inverse. A library declaring more than one of these operations must state their separate types and laws and establish any claimed relationship between them. This keeps a transpose used in differentiation from silently acquiring a recovery or resource-cancellation contract during lowering.

\subsection*{9.2 Realizing the contract in ordinary MLIR}\label{realizing-the-contract-in-ordinary-mlir}
\addcontentsline{toc}{subsection}{9.2 Realizing the contract in ordinary MLIR}

Our proposed PSG interface retains the paired operation identities, value indices, resource-use conditions and admitted proof rules. Baker establishes or dispatches their premises, then selects a realization whose operations satisfy the source contract. Alex witnesses that settled construction using ordinary operations for values, control flow, memory and calls. Composer\textquotesingle s backend maps them to the selected hardware.

A small admission example separates those duties. Suppose a library\textquotesingle s matched values range over the unsigned integers \(0\) through \(255\), its equality is exact value equality under the canonical injective eight-bit encoding, and its contract has explicit success and mismatch continuations. Let \(v\) be the required value and \(x\) the supplied value. For already obtained scalar operands, an illustrative realization of the match decision is

\begin{lstlisting}
%matches = arith.cmpi eq, %supplied, %required : i8
cf.cond_br %matches, ^matched, ^mismatch
\end{lstlisting}

This is a block fragment. The surrounding realization supplies the two continuations and their resource transitions. The preservation proof relates \lstinline!%supplied! to \(x\) and \lstinline!%required! to \(v\). Injectivity gives \(\operatorname{enc}(x)=\operatorname{enc}(v)\iff x=v\), so the selected branch agrees with the source equality judgment. The surrounding continuations have admitted contracts for the permitted consumption or transfer of the paired resource on each branch. A value index that is known statically can use a constant or a proved comparison result; a dynamic index requires the corresponding value information at the decision.

If the premises establish \(x=v\), a checked rewrite can select the success path directly while preserving the required demand, effects and resource use. If the match remains dynamic, the emitted comparison preserves its specified mismatch behavior. Thus the proof can justify removing a check or retaining it, according to the same source contract. The erased type need not occupy a runtime tag or box once its remaining duties are represented by the operations and their compiler evidence.

\includegraphics[width=1\linewidth,height=\textheight,keepaspectratio]{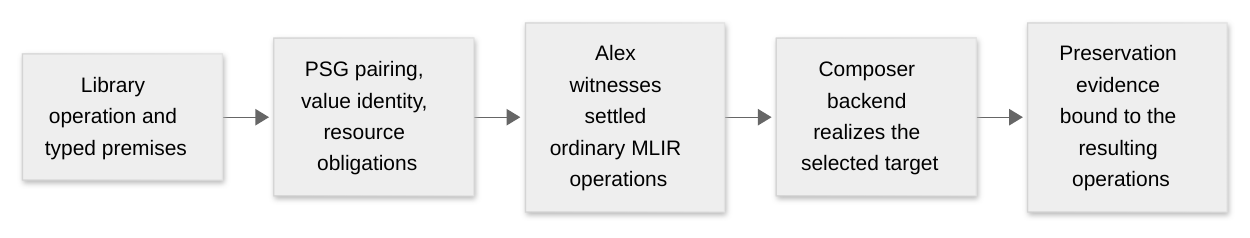}

A reverse operation follows the same discipline. Its selected inverse law determines the admitted state domain and any retained control information. Exact reconstruction, bounded approximation and recorded replay have different contracts. Each lowering preserves the one exposed by its library interface. Physical quantum operations additionally carry the device interpretation and error assumptions relevant to their promised behavior.

This division keeps the middle end thin while permitting a richer source language. MLIR carries the operations implementing the construction, and our graph and proof interfaces carry the reason those operations satisfy it. Reusable realization rules can then deliver an unboxed computation graph to the target without requiring the application developer to restate the type\textquotesingle s categorical account. The companion NFT paper develops the utility of these types on that explicit admission boundary.

\section*{10. Differential Compilation and Evidence Reuse}\label{differential-compilation-and-evidence-reuse}
\addcontentsline{toc}{section}{10. Differential Compilation and Evidence Reuse}

A developer edits a program more often than they replace its entire design. Our preservation account should retain useful work across those edits. The relevant question is which conclusions still have valid support, including the evidence linking them to the operations already compiled.

\subsection*{10.1 Re-establishing support after an edit}\label{re-establishing-support-after-an-edit}
\addcontentsline{toc}{subsection}{10.1 Re-establishing support after an edit}

An analysis fixed point can be internally consistent and still describe an earlier program. Suppose two cached facts cite one another after the source premise that supported them has been withdrawn. Their agreement is insufficient for reuse. Re-settlement follows current derivations and retracts facts whose required support has changed. Inductive derivations retain their well-founded justification; cyclic or coinductive evidence uses the admission rule of its particular proof system.

Our specification makes the affected region depend on semantic support: actual participants, type and value premises, collection membership where consumed, rule identities, target facts, resource and boundary contracts. A changed source location is the starting point for this closure. A specialized or inlined consumer may depend on more than a published signature. When precise support is unavailable, reconsidering the owning region provides the conservative scope.

For a geometric consumer, the closure also includes trace construction, its observation map, density and normalization rules, coordinate scales, metric weights and selected tangent directions. Changing any of these can invalidate metric entries, kernel or conditioning evidence, and the optimizer step that consumes them, even if an output-law equivalence remains valid. Cached estimates additionally retain the parameter state and sampling assumptions under which they were obtained; reuse across states requires its own accepted rule. A change confined to an independent region can preserve those instances whose support and correspondence remain established. The existing dependency mechanism remains the owner of these geometric facts.

Our first-release architecture uses that closure for differential compilation. CCS/Baker re-establishes the affected facts and settled interfaces. Authorized witness scopes identify the current graph occurrences to re-lower. Alex witnesses those scopes, and the selected backend realizes their changes. Valid fragments and evidence outside the affected closure remain eligible for reuse. Publication binds each replacement to its accepted revision and receiving contracts.

\subsection*{10.2 The FFI example across revisions}\label{the-ffi-example-across-revisions}
\addcontentsline{toc}{subsection}{10.2 The FFI example across revisions}

Return to the call in §8. A change from the source range \([0,1023]\) to a larger admitted range can change its storage and argument conversion. A changed foreign declaration affects ABI coverage and call realization. A changed receiver affects the result conversion. Moving a callback\textquotesingle s environment can change lifetime support even when the function signature stays identical.

Each change has a different dependency footprint. The compiler follows the premises actually consumed by the boundary rather than rebuilding every proof about unrelated program operations. A still-valid dimensional derivation can support a revised range or relational argument through its mode interface. A checked foreign-call rule can be reused with new parameters and new premise evidence. This is cumulative proof construction with explicit revalidation at the changed application.

The quantitative example has the same behavior. Changing the pivoting rule or sampler changes the premises of the Urschel theorem instance. Changing the target precision affects the arithmetic bridge and conversion contract. Matrix dimensions and structural relationships whose support remains valid can be retained. This lets mathematical evidence participate in ordinary dependency-directed compilation.

\subsection*{10.3 Recombination and cost}\label{recombination-and-cost}
\addcontentsline{toc}{subsection}{10.3 Recombination and cost}

Reusing fragments also requires compatible boundaries. The join-tree rule of §2.3 supplies one sufficient relational assembly condition. Other cases use their admitted joint contract, with tightly coupled components settled together. Recombination preserves the relationships between the source participants, proof instances and resulting operations even when a rewrite splits or merges a previous segment.

The resulting efficiency claim is specific. A retained derivation avoids reconstructing that derivation; a retained realized fragment avoids repeating its lowering. Support validation, invalidation, changed-region analysis and boundary checks remain part of the work. The affected region may expand when dependencies cross a nominal segment boundary. The benefit comes from using evidence to determine which work is necessary for this revision.

This is the compiler-level counterpart of justified erasure. At runtime, established structure permits efficient unboxed operations. During development, that same structure permits reuse of unaffected analyses, proofs and realizations. Formalization supplies information with which the compiler can remove repeated work across both phases.

\section*{11. A Thin Middle End with Retained Integrity}\label{a-thin-middle-end-with-retained-integrity}
\addcontentsline{toc}{section}{11. A Thin Middle End with Retained Integrity}

Our construction builds on practical compiler infrastructure. Gabriel\textquotesingle s {\textquotedblleft{}Worse} is {Better\textquotedblright{}} frames the recurring tension between a complete design and an adoptable implementation \hyperref[ref-gabriel]{{[}31{]}}. We have described our use of MLIR as {\textquotedblleft{}New} Jersey wheels under an MIT {frame\textquotedblright{}.} A substantial existing substrate serves a language whose source contracts determine what that substrate must preserve. Scheme\textquotesingle s nanopass methodology supplies a related discipline of small transformations with explicit responsibilities \hyperref[ref-sarkar]{{[}32{]}}.

MLIR provides nested regions, operation structure and extensible back-end dialects across heterogeneous targets. The considerable community investment in MLIR for first-class verification dialects \hyperref[ref-fehr]{{[}33{]}} and verified SSA rewriting \hyperref[ref-bhat]{{[}34{]}} demonstrates ways of integrating formal reasoning with that infrastructure. Our source-directed account places semantic decisions and their joint proof dependencies in the PSG, uses Alex (our Library of Alexandria) middle end component to witness their settled consequences, and binds the selected backend realization to those contracts. This division makes our "thin middle end" of the Composer compiler an active part of the integrity chain.

The foundations contribute at different points. Parametricity supplies reusable facts for a defined interpretation. Ohori\textquotesingle s proof transformations explain how matching local contracts can compose through compilation. Hăvărneanu\textquotesingle s shift-adjunction results inform interfaces between reasoning disciplines. Urschel supplies a recent example of quantitative evidence that can enter those interfaces with explicit assumptions and bounds. Our compiler account connects these results to the actual operations, representations and resources that consume them.

The FFI work substantiates that connection in a concrete implementation path. Argument coverage, return conversion, callable formation and evidence identity are retained across source analysis and witnessing. The negative and fractional type case develops the corresponding admission contract for richer value and resource relationships. Both illustrate why an ordinary target operation can carry the consequence of a stronger source proof when its semantic correspondence remains available through lowering.

For application developers, the result is an integrated compilation discipline: supported proofs arise from program structure and reusable libraries, native realizations preserve the contracts they consume, and differential compilation retains work whose support survives an edit. Integrity and efficiency depend on the same information. Our fixed-point scaffolding organizes its passage from the source program to the unboxed computation that the selected hardware receives.

\begin{center}\rule{0.5\linewidth}{0.5pt}\end{center}

\section*{Appendix. Inspectable examples at the witness boundary}\label{appendix.-inspectable-examples-at-the-witness-boundary}
\addcontentsline{toc}{section}{Appendix. Inspectable examples at the witness boundary}

The first two examples use source programs from the dimensional regression module. They expose facts that must survive checking and elaboration. The third is a worked arithmetic validation obligation. Together with the FFI and pairing examples in §§8--9, they identify the source-to-target relationships that a pass theorem or per-instance validation must establish.

\subsection*{A. A measured identity survives elaboration}\label{a.-a-measured-identity-survives-elaboration}
\addcontentsline{toc}{subsection}{A. A measured identity survives elaboration}

The regression named \lstinline!power survives elaboration! checks this function, in a module declaring measure \lstinline!m!:

\begin{lstlisting}
let keep (x: float<m^2>) = x
\end{lstlisting}

It inspects the resulting binding type and verifies that the argument retains \(m^2\), including the distinction from \(m\). Separate identity cases reject unification of meter and second quantities and distinguish qualified base-measure identities. These checks exercise the type representation and elaboration path where measure syntax could otherwise be discarded.

If a platform and the applicable range and representation contract select binary64 for a concrete instance, its runtime identity operation can have this form:

\begin{lstlisting}
func.func @keep(%arg0: f64) -> f64 {
  func.return %arg0 : f64
}
\end{lstlisting}

The preservation relation associates this argument and result with the checked \(m^2\) source identity. Erasing the explicit measure representation is permissible after its relevant typing and boundary obligations are retained or discharged. The elaboration regression protects that source identity. A validation or pass theorem connects it to the lowered argument and result. Together they account for the unboxed \lstinline!f64! realization.

\subsection*{B. A generic record retains dimensions while layout remains deferred}\label{b.-a-generic-record-retains-dimensions-while-layout-remains-deferred}
\addcontentsline{toc}{subsection}{B. A generic record retains dimensions while layout remains deferred}

The regression named \lstinline!layout deferral keeps generic measured code usable! contains:

\begin{lstlisting}
type Reading<[<Measure>] 'u, 'a> =
    { Context: 'a; Quantity: float<'u>; Valid: bool }
let reading = { Context = true; Quantity = 1.0<m>; Valid = true }
let length = reading.Quantity
\end{lstlisting}

The test checks that \lstinline!length! has the meter-valued type and that the generic declaration has not acquired a concrete layout before its parameters are known. The measure parameter participates in source identity even though it requires no numeric payload slot. The ordinary parameter can affect the record\textquotesingle s size and alignment when instantiated.

A native instance needs a layout derived from its instantiated fields and selected target. BAREWire\textquotesingle s mapping and the primitive access rules must agree on those fields, including representation, offset, extent, and alignment. An unresolved layout must remain unresolved until those facts exist. A later known field cannot repair an earlier unknown field by causing its size to be ignored.

The lifetime of an instance is another fact. A record with a finite number of fields may contain a reference to dynamically sized storage. A local binding does not prove that the referenced storage is local, and copying an immutable reference does not duplicate the object it denotes. These are the distinctions a placement proof must preserve before an allocation witness selects its realization.

\subsection*{C. A signed-width obligation}\label{c.-a-signed-width-obligation}
\addcontentsline{toc}{subsection}{C. A signed-width obligation}

Consider a phase counter with an established invariant \(0\le p\le1023\) and transition

\[p'=(p+1)\bmod1024.\]

Under a signed two\textquotesingle s-complement representation policy, the state needs eleven bits: its maximum is \(2^{10}-1=1023\). The increment may equal \(1024\), so twelve signed bits suffice for the intermediate. An unsigned policy could represent the state in ten bits, with different adaptation rules. Width and signedness are therefore part of the representation contract.

For the signed policy, a lowering can sign-extend the state from eleven to twelve bits, add one, take the remainder, and narrow the result to eleven bits. The following QF\_BV obligation checks that narrowing by a round trip. Nonnegative twelve-bit operands permit unsigned remainder here, and their declared range ensures the addition does not overflow the signed intermediate.

\begin{lstlisting}
(set-logic QF_BV)
(declare-fun p () (_ BitVec 12))
(assert (bvsge p (_ bv0 12)))
(assert (bvslt p (_ bv1024 12)))
(define-fun inc () (_ BitVec 12) (bvadd p (_ bv1 12)))
(define-fun wrap () (_ BitVec 12) (bvurem inc (_ bv1024 12)))
(define-fun narrowed () (_ BitVec 11) ((_ extract 10 0) wrap))
(assert (not (= ((_ sign_extend 1) narrowed) wrap)))
(check-sat)
\end{lstlisting}

The query is unsatisfiable: no counterexample loses the signed value through this narrowing. Narrowing to ten signed bits instead admits a counterexample, such as \(p=511\), for which the next value \(512\) would be read back as \(-512\). Checking only that the result is below \(2048\) would miss that signedness error.

MLIR\textquotesingle s SMT dialect can represent this obligation alongside an operation-level verification path \hyperref[ref-mlir]{{[}28{]}}. The correspondence check binds the query to the actual addition, remainder and truncation, and to evidence establishing the input invariant. An \lstinline!unsat! result then establishes the encoded narrowing statement under those premises. The compiler carries that result with the operation it justifies.

\begin{center}\rule{0.5\linewidth}{0.5pt}\end{center}

\section*{Software Availability}\label{software-availability}
\addcontentsline{toc}{section}{Software Availability}

The Clef language, Composer compiler, and supporting libraries described in this paper are developed under the Fidelity Framework project. Source repositories are available at \url{https://github.com/FidelityFramework}. The language specification, design rationale, and compiler documentation are published at \url{https://clef-lang.com}. Central components of the framework are dual-licensed; terms are detailed in each repository. The compiler design connects these components through the obligation and preservation interfaces described in this paper.

\section*{References}\label{references}
\addcontentsline{toc}{section}{References}

\leavevmode\protect\phantomsection\label{ref-ohori}{}{[}1{]} Ohori, A. (2007). A proof theory for machine code. \emph{ACM Transactions on Programming Languages and Systems}, 29(6), Article 36. \href{https://doi.org/10.1145/1286821.1286827}{DOI:10.1145/1286821.1286827}.

\leavevmode\protect\phantomsection\label{ref-reynolds}{}{[}2{]} Reynolds, J. C. (1983). Types, abstraction and parametric polymorphism. In \emph{Information Processing 83}.

\leavevmode\protect\phantomsection\label{ref-wadler}{}{[}3{]} Wadler, P. (1989). Theorems for free! In \emph{FPCA}, 347--359. \href{https://doi.org/10.1145/99370.99404}{DOI:10.1145/99370.99404}.

\leavevmode\protect\phantomsection\label{ref-kennedy1996}{}{[}4{]} Kennedy, A. (1996). \emph{Programming Languages and Dimensions}. PhD thesis, University of Cambridge, Technical Report 391. \href{https://www.cl.cam.ac.uk/techreports/UCAM-CL-TR-391.html}{Report}.

\leavevmode\protect\phantomsection\label{ref-kennedy2009}{}{[}5{]} Kennedy, A. (2009). Types for units-of-measure: Theory and practice. In \emph{Central European Functional Programming School}.

\leavevmode\protect\phantomsection\label{ref-pruiksma}{}{[}6{]} Pruiksma, K., Chargin, W., Pfenning, F., and Reed, J. (2018). \emph{Adjoint Logic}. Manuscript.

\leavevmode\protect\phantomsection\label{ref-havarneanu}{}{[}7{]} Hăvărneanu, A. (2026). \emph{Classical Adjoint Logic}. Pre-publication manuscript, July 12.

\leavevmode\protect\phantomsection\label{ref-urschel}{}{[}8{]} Urschel, J. (2026). On the growth factor of random matrices. \href{https://arxiv.org/abs/2610.06785v1}{arXiv:2610.06785v1}.

\leavevmode\protect\phantomsection\label{ref-dts}{}{[}9{]} Haynes, H. (2026). Dimensional type systems and deterministic memory management: Design-time semantic preservation in native compilation. \href{https://arxiv.org/abs/2603.16437}{arXiv:2603.16437}.

\leavevmode\protect\phantomsection\label{ref-phg}{}{[}10{]} Haynes, H. (2026). The program hypergraph: Multi-way relational structure for geometric algebra, spatial compute, and physics-aware compilation. \href{https://arxiv.org/abs/2603.17627}{arXiv:2603.17627}.

\leavevmode\protect\phantomsection\label{ref-dbc}{}{[}11{]} Haynes, H. (2026). Decidable by construction: Design-time verification for truly fearless systems. \href{https://arxiv.org/abs/2603.25414}{arXiv:2603.25414}.

\leavevmode\protect\phantomsection\label{ref-tarau}{}{[}12{]} Tarau, P. (2008). Isomorphic data encodings and their generalization to hylomorphisms on hereditarily finite data types. \href{https://arxiv.org/abs/0808.2953}{arXiv:0808.2953}.

\leavevmode\protect\phantomsection\label{ref-beeri}{}{[}13{]} Beeri, C., Fagin, R., Maier, D., and Yannakakis, M. (1983). On the desirability of acyclic database schemes. \emph{Journal of the ACM}, 30(3), 479--513. \href{https://doi.org/10.1145/2402.322389}{DOI:10.1145/2402.322389}.

\leavevmode\protect\phantomsection\label{ref-iris}{}{[}14{]} Jung, R., Krebbers, R., Jourdan, J.-H., Bizjak, A., Birkedal, L., and Dreyer, D. (2018). Iris from the ground up: A modular foundation for higher-order concurrent separation logic. \emph{Journal of Functional Programming}, 28, e20. \href{https://www.mpi-sws.org/~dreyer/papers/iris-ground-up/paper.pdf}{Paper}.

\leavevmode\protect\phantomsection\label{ref-actris}{}{[}15{]} Hinrichsen, J. K., Krebbers, R., and Bengtson, J. (2022). Actris 2.0: Asynchronous session-type based reasoning in separation logic. \emph{Logical Methods in Computer Science}. \href{https://iris-project.org/actris/}{Paper and mechanization}.

\leavevmode\protect\phantomsection\label{ref-aneris}{}{[}16{]} Aneris project. \emph{Program logic for developing and verifying distributed systems}. \href{https://github.com/logsem/aneris}{Development and documentation}.

\leavevmode\protect\phantomsection\label{ref-rclib}{}{[}17{]} Gondelman, L., Hinrichsen, J. K., Pereira, M., Timany, A., and Birkedal, L. (2023). Verifying reliable network components in a distributed separation logic with dependent separation protocols. \emph{Proceedings of the ACM on Programming Languages}, ICFP. \href{https://iris-project.org/pdfs/2023-icfp-relcom.pdf}{Paper}.

\leavevmode\protect\phantomsection\label{ref-barthe}{}{[}18{]} Barthe, G., Espitau, T., Grégoire, B., Hsu, J., Stefanesco, L., and Strub, P.-Y. (2015). Relational reasoning via probabilistic coupling. \href{https://arxiv.org/abs/1509.03476}{arXiv:1509.03476}.

\leavevmode\protect\phantomsection\label{ref-jasmin}{}{[}19{]} Arranz-Olmos, S., Barthe, G., Blatter, L., Grégoire, B., Laporte, V., and Torrini, P. (2025). The Jasmin compiler preserves cryptographic security. \href{https://arxiv.org/abs/2511.11292v1}{arXiv:2511.11292v1}. Revised as \emph{KEM-IND-CCA-Preserving Compilation of Jasmin\textquotesingle s ML-KEM} (2026), \href{https://arxiv.org/abs/2511.11292v2}{arXiv:2511.11292v2}.

\leavevmode\protect\phantomsection\label{ref-adm}{}{[}20{]} Haynes, H. (2026). Adaptive domain models: Bayesian evolution, warm rotation, and principled training for geometric and neuromorphic AI. \href{https://arxiv.org/abs/2603.18104}{arXiv:2603.18104}.

\leavevmode\protect\phantomsection\label{ref-cvc5}{}{[}21{]} cvc5 developers. Solver outputs: Models, cores and proofs. \emph{cvc5 tutorials}. \href{https://cvc5.github.io/tutorials/beginners/outputs.html}{Documentation}.

\leavevmode\protect\phantomsection\label{ref-huet}{}{[}22{]} Huet, G. (1997). The zipper. \emph{Journal of Functional Programming}, 7(5), 549--554. \href{https://doi.org/10.1017/S0956796897002864}{DOI:10.1017/S0956796897002864}.

\leavevmode\protect\phantomsection\label{ref-karwowski-geometry-2026}{}{[}23{]} Karwowski, J., Nielsen, F., and Staton, S. (2026). Compositional information geometry of probabilistic programs. \href{https://arxiv.org/abs/2610.17886v1}{arXiv:2610.17886v1}.

\leavevmode\protect\phantomsection\label{ref-phase}{}{[}24{]} Harper, R., Mitchell, J. C., and Moggi, E. (1990). Higher-order modules and the phase distinction. In \emph{POPL}.

\leavevmode\protect\phantomsection\label{ref-eades}{}{[}25{]} Eades, H., and Orchard, D. (2020). Grading adjoint logic. \href{https://arxiv.org/abs/2006.08854}{arXiv:2006.08854}.

\leavevmode\protect\phantomsection\label{ref-qkb}{}{[}26{]} Qian, Z., Kavvos, G. A., and Birkedal, L. (2021). Client-server sessions in linear logic. In \emph{ICFP}. \href{https://arxiv.org/abs/2010.13926}{arXiv:2010.13926}.

\leavevmode\protect\phantomsection\label{ref-johann}{}{[}27{]} Johann, P., and Voigtländer, J. (2004). Free theorems in the presence of seq. In \emph{POPL}, 99--110. \href{https://doi.org/10.1145/964001.964010}{DOI:10.1145/964001.964010}.

\leavevmode\protect\phantomsection\label{ref-mlir}{}{[}28{]} LLVM Project. \emph{MLIR dialect documentation}: \href{https://mlir.llvm.org/docs/Dialects/ArithOps/}{Arith}, \href{https://mlir.llvm.org/docs/Dialects/Func/}{Func}, \href{https://mlir.llvm.org/docs/Dialects/ControlFlowDialect/}{Control Flow}, and \href{https://mlir.llvm.org/docs/Dialects/SMT/}{SMT}.

\leavevmode\protect\phantomsection\label{ref-nft}{}{[}29{]} Haynes, H. (2026). Negative and fractional types in the Fidelity Framework: A proposed operational core for reversibility and resource pairing. \href{https://arxiv.org/abs/2606.04352}{arXiv:2606.04352}.

\leavevmode\protect\phantomsection\label{ref-chen}{}{[}30{]} Chen, C., and Sabry, A. (2021). A computational interpretation of compact closed categories: Reversible programming with negative and fractional types. \emph{Proceedings of the ACM on Programming Languages}, 5(POPL), Article 9.

\leavevmode\protect\phantomsection\label{ref-gabriel}{}{[}31{]} Gabriel, R. P. (1991). Lisp: Good news, bad news, how to win big. Includes \href{https://dreamsongs.com/WIB.html}{{\textquotedblleft{}Worse} is {Better\textquotedblright{}}}.

\leavevmode\protect\phantomsection\label{ref-sarkar}{}{[}32{]} Sarkar, D., Waddell, O., and Dybvig, R. K. (2004). A nanopass infrastructure for compiler education. In \emph{ICFP}.

\leavevmode\protect\phantomsection\label{ref-fehr}{}{[}33{]} Fehr, M., Fan, Y., Pompougnac, H., Regehr, J., and Grosser, T. (2025). First-class verification dialects for MLIR. \emph{Proceedings of the ACM on Programming Languages}, 9(PLDI), Article 206.

\leavevmode\protect\phantomsection\label{ref-bhat}{}{[}34{]} Bhat, S., Keizer, A., Hughes, C., Goens, A., and Grosser, T. (2024). Verifying peephole rewriting in SSA compiler IRs. In \emph{ITP}.
\end{document}